\documentclass[aps,prl,twocolumn,showpacs,groupedaddress]{revtex4}  
\usepackage{graphicx}  
\usepackage{dcolumn}   
\usepackage{bm}        
\usepackage{amssymb}   
\begin{document}
\def\slashchar#1{\setbox0=\hbox{$#1$}           
   \dimen0=\wd0                                 
   \setbox1=\hbox{/} \dimen1=\wd1               
   \ifdim\dimen0>\dimen1                        
      \rlap{\hbox to \dimen0{\hfil/\hfil}}      
      #1                                        
   \else                                        
      \rlap{\hbox to \dimen1{\hfil$#1$\hfil}}   
      /                                         
   \fi}                                         %
\def\etmiss{\slashchar{E}_T}			%
\def\htmiss{\slashchar{H}_T}			%


\hspace{5.2in} \mbox{Fermilab-Pub-07/113-E}

\title{Search for third-generation leptoquarks in $p\bar{p}$ 
collisions at $\sqrt{s}$ = 1.96 TeV}
%
%
\author{                                                                      
V.M.~Abazov,$^{35}$                                                           
B.~Abbott,$^{75}$                                                             
M.~Abolins,$^{65}$                                                            
B.S.~Acharya,$^{28}$                                                          
M.~Adams,$^{51}$                                                              
T.~Adams,$^{49}$                                                              
E.~Aguilo,$^{5}$                                                              
S.H.~Ahn,$^{30}$                                                              
M.~Ahsan,$^{59}$                                                              
G.D.~Alexeev,$^{35}$
% ...
\\                                                                            
\vskip 0.30cm                                                                 
\centerline{(D\O\ Collaboration)}                                             
\vskip 0.30cm                                                                 
}                                                                             
\affiliation{                                                                 
\centerline{$^{1}$Universidad de Buenos Aires, Buenos Aires, Argentina}       
\centerline{$^{2}$LAFEX, Centro Brasileiro de Pesquisas F{\'\i}sicas,         
                  Rio de Janeiro, Brazil}                                     
\centerline{$^{3}$Universidade do Estado do Rio de Janeiro,                   
                  Rio de Janeiro, Brazil}                                     
\centerline{$^{4}$Instituto de F\'{\i}sica Te\'orica, Universidade            
                  Estadual Paulista, S\~ao Paulo, Brazil}                     
\centerline{$^{5}$University of Alberta, Edmonton, Alberta, Canada,           
                  Simon Fraser University, Burnaby, British Columbia, Canada,}
\centerline{York University, Toronto, Ontario, Canada, and                    
                  McGill University, Montreal, Quebec, Canada}                
\centerline{$^{6}$University of Science and Technology of China, Hefei,       
                  People's Republic of China}                                 
\centerline{$^{7}$Universidad de los Andes, Bogot\'{a}, Colombia}             
\centerline{$^{8}$Center for Particle Physics, Charles University,            
                  Prague, Czech Republic}                                     
\centerline{$^{9}$Czech Technical University, Prague, Czech Republic}         
\centerline{$^{10}$Center for Particle Physics, Institute of Physics,         
                   Academy of Sciences of the Czech Republic,                 
                   Prague, Czech Republic}                                    
\centerline{$^{11}$Universidad San Francisco de Quito, Quito, Ecuador}        
\centerline{$^{12}$Laboratoire de Physique Corpusculaire, IN2P3-CNRS,         
                   Universit\'e Blaise Pascal, Clermont-Ferrand, France}      
\centerline{$^{13}$Laboratoire de Physique Subatomique et de Cosmologie,      
                   IN2P3-CNRS, Universite de Grenoble 1, Grenoble, France}    
\centerline{$^{14}$CPPM, IN2P3-CNRS, Universit\'e de la M\'editerran\'ee,     
                   Marseille, France}                                         
\centerline{$^{15}$Laboratoire de l'Acc\'el\'erateur Lin\'eaire,              
                   IN2P3-CNRS et Universit\'e Paris-Sud, Orsay, France}       
\centerline{$^{16}$LPNHE, IN2P3-CNRS, Universit\'es Paris VI and VII,         
                   Paris, France}                                             
\centerline{$^{17}$DAPNIA/Service de Physique des Particules, CEA, Saclay,    
                   France}                                                    
\centerline{$^{18}$IPHC, Universit\'e Louis Pasteur et Universit\'e           
                   de Haute Alsace, CNRS, IN2P3, Strasbourg, France}          
\centerline{$^{19}$IPNL, Universit\'e Lyon 1, CNRS/IN2P3, Villeurbanne, France
                   and Universit\'e de Lyon, Lyon, France}                    
\centerline{$^{20}$III. Physikalisches Institut A, RWTH Aachen,               
                   Aachen, Germany}                                           
\centerline{$^{21}$Physikalisches Institut, Universit{\"a}t Bonn,             
                   Bonn, Germany}                                             
\centerline{$^{22}$Physikalisches Institut, Universit{\"a}t Freiburg,         
                   Freiburg, Germany}                                         
\centerline{$^{23}$Institut f{\"u}r Physik, Universit{\"a}t Mainz,            
                   Mainz, Germany}                                            
\centerline{$^{24}$Ludwig-Maximilians-Universit{\"a}t M{\"u}nchen,            
                   M{\"u}nchen, Germany}                                      
\centerline{$^{25}$Fachbereich Physik, University of Wuppertal,               
                   Wuppertal, Germany}                                        
\centerline{$^{26}$Panjab University, Chandigarh, India}                      
\centerline{$^{27}$Delhi University, Delhi, India}                            
\centerline{$^{28}$Tata Institute of Fundamental Research, Mumbai, India}     
\centerline{$^{29}$University College Dublin, Dublin, Ireland}                
\centerline{$^{30}$Korea Detector Laboratory, Korea University,               
                   Seoul, Korea}                                              
\centerline{$^{31}$SungKyunKwan University, Suwon, Korea}                     
\centerline{$^{32}$CINVESTAV, Mexico City, Mexico}                            
\centerline{$^{33}$FOM-Institute NIKHEF and University of                     
                   Amsterdam/NIKHEF, Amsterdam, The Netherlands}              
\centerline{$^{34}$Radboud University Nijmegen/NIKHEF, Nijmegen, The          
                  Netherlands}                                                
\centerline{$^{35}$Joint Institute for Nuclear Research, Dubna, Russia}       
\centerline{$^{36}$Institute for Theoretical and Experimental Physics,        
                   Moscow, Russia}                                            
\centerline{$^{37}$Moscow State University, Moscow, Russia}                   
\centerline{$^{38}$Institute for High Energy Physics, Protvino, Russia}       
\centerline{$^{39}$Petersburg Nuclear Physics Institute,                      
                   St. Petersburg, Russia}                                    
\centerline{$^{40}$Lund University, Lund, Sweden, Royal Institute of          
                   Technology and Stockholm University, Stockholm,            
                   Sweden, and}                                               
\centerline{Uppsala University, Uppsala, Sweden}                              
\centerline{$^{41}$Physik Institut der Universit{\"a}t Z{\"u}rich,            
                   Z{\"u}rich, Switzerland}                                   
\centerline{$^{42}$Lancaster University, Lancaster, United Kingdom}           
\centerline{$^{43}$Imperial College, London, United Kingdom}                  
\centerline{$^{44}$University of Manchester, Manchester, United Kingdom}      
\centerline{$^{45}$University of Arizona, Tucson, Arizona 85721, USA}         
\centerline{$^{46}$Lawrence Berkeley National Laboratory and University of    
                   California, Berkeley, California 94720, USA}               
\centerline{$^{47}$California State University, Fresno, California 93740, USA}
\centerline{$^{48}$University of California, Riverside, California 92521, USA}
\centerline{$^{49}$Florida State University, Tallahassee, Florida 32306, USA} 
\centerline{$^{50}$Fermi National Accelerator Laboratory,                     
            Batavia, Illinois 60510, USA}                                     
\centerline{$^{51}$University of Illinois at Chicago,                         
            Chicago, Illinois 60607, USA}                                     
\centerline{$^{52}$Northern Illinois University, DeKalb, Illinois 60115, USA} 
\centerline{$^{53}$Northwestern University, Evanston, Illinois 60208, USA}    
\centerline{$^{54}$Indiana University, Bloomington, Indiana 47405, USA}       
\centerline{$^{55}$University of Notre Dame, Notre Dame, Indiana 46556, USA}  
\centerline{$^{56}$Purdue University Calumet, Hammond, Indiana 46323, USA}    
\centerline{$^{57}$Iowa State University, Ames, Iowa 50011, USA}              
\centerline{$^{58}$University of Kansas, Lawrence, Kansas 66045, USA}         
\centerline{$^{59}$Kansas State University, Manhattan, Kansas 66506, USA}     
\centerline{$^{60}$Louisiana Tech University, Ruston, Louisiana 71272, USA}   
\centerline{$^{61}$University of Maryland, College Park, Maryland 20742, USA} 
\centerline{$^{62}$Boston University, Boston, Massachusetts 02215, USA}       
\centerline{$^{63}$Northeastern University, Boston, Massachusetts 02115, USA} 
\centerline{$^{64}$University of Michigan, Ann Arbor, Michigan 48109, USA}    
\centerline{$^{65}$Michigan State University,                                 
            East Lansing, Michigan 48824, USA}                                
\centerline{$^{66}$University of Mississippi,                                 
            University, Mississippi 38677, USA}                               
\centerline{$^{67}$University of Nebraska, Lincoln, Nebraska 68588, USA}      
\centerline{$^{68}$Princeton University, Princeton, New Jersey 08544, USA}    
\centerline{$^{69}$State University of New York, Buffalo, New York 14260, USA}
\centerline{$^{70}$Columbia University, New York, New York 10027, USA}        
\centerline{$^{71}$University of Rochester, Rochester, New York 14627, USA}   
\centerline{$^{72}$State University of New York,                              
            Stony Brook, New York 11794, USA}                                 
\centerline{$^{73}$Brookhaven National Laboratory, Upton, New York 11973, USA}
\centerline{$^{74}$Langston University, Langston, Oklahoma 73050, USA}        
\centerline{$^{75}$University of Oklahoma, Norman, Oklahoma 73019, USA}       
\centerline{$^{76}$Oklahoma State University, Stillwater, Oklahoma 74078, USA}
\centerline{$^{77}$Brown University, Providence, Rhode Island 02912, USA}     
\centerline{$^{78}$University of Texas, Arlington, Texas 76019, USA}          
\centerline{$^{79}$Southern Methodist University, Dallas, Texas 75275, USA}   
\centerline{$^{80}$Rice University, Houston, Texas 77005, USA}                
\centerline{$^{81}$University of Virginia, Charlottesville,                   
            Virginia 22901, USA}                                              
\centerline{$^{82}$University of Washington, Seattle, Washington 98195, USA}  
}                                                                             

\date{May 6, 2007}

\begin{abstract}
We report on a search for charge-1/3 third-generation leptoquarks 
(LQ) produced in $p\bar{p}$ collisions at $\sqrt{s}=1.96$~TeV using 
the D0 detector at Fermilab. Third generation leptoquarks
are assumed to be produced in pairs and to decay to a tau neutrino and a 
$b$ quark with branching fraction~$B$. We place upper limits on 
$\sigma(p\bar{p} \rightarrow LQ\overline{LQ})\times B^2$ as a function
of the leptoquark mass $M_{LQ}$. 
Assuming $B$ = 1, we exclude at the 95\% confidence level third-generation 
scalar leptoquarks with $M_{LQ} < 229$~GeV.
\end{abstract}

\maketitle
Leptoquarks (LQ) are bosons predicted in many extensions of the standard 
model (SM) \cite{theory1}. They carry both nonzero lepton and color quantum 
numbers and decay to a lepton and quark (or antiquark). 
To satisfy experimental limits on lepton number violation,
on flavor-changing neutral currents, and on proton decay, leptoquarks 
of mass accessible to current collider experiments are constrained to couple
to only one generation of leptons and quarks \cite{theory2}.
Therefore, only leptoquarks that couple within a single generation are
considered here.

This Letter reports the results of a search for charge-1/3 
third-generation leptoquarks produced in $p\bar{p}$ collisions
at $\sqrt{s}=1.96$ TeV. We assume that leptoquarks are produced in pairs by 
$q\bar{q}$ annihilation or $gg$ fusion, i.e., 
$p+\bar{p} \rightarrow LQ+\overline{LQ}+X$. 
These processes are independent of the unknown leptoquark-lepton-quark coupling,
and the pair production cross section has been calculated including 
next-to-leading order terms for scalar leptoquarks \cite{Kramer:1997hh}.
Such leptoquarks would decay into either a $\nu_\tau$  
plus a $b$ quark or a $\tau$ lepton plus a $t$ quark. We search for the 
decay signature where both leptoquarks 
decay via $ LQ \rightarrow \nu_\tau + b$ with branching 
fraction $B$, resulting in a $\nu_\tau\bar{\nu}_\tau b\bar{b}$ final state. 
Upper limits on the cross section times $B^2$ as a function of leptoquark mass 
($M_{LQ}$) are measured and then used to determine lower limits on $M_{LQ}$ 
assuming they are scalar for which the calculated cross section is lower and 
better determined than that for vector leptoquarks which have only 
been calculated to leading order \cite{VLQ-xsec}.
Previous limits from Fermilab Run I data were reported by both the D0
\cite{D0_lq3bnu:1998} and CDF \cite{CDFlq3_taub:1997,CDF_lq3bnu:2000}
collaborations based on significantly smaller integrated luminosities
and at a slightly lower center-of-mass energy compared with the Run II 
data available now.

The upgraded Run II D0 detector \cite{d0nim} consists 
of layered systems surrounding the interaction point. Closest to the beam 
are the silicon microstrip tracker and a central fiber tracker, both 
immersed in the field of a 2~T solenoid. These measure the momenta of charged
particles and reconstruct primary and secondary vertices. Jets and electrons 
are reconstructed using the pattern of energy deposited in three 
uranium/liquid-argon calorimeters outside the tracking system with a central 
section covering $|\eta|< 1.1$ and two end calorimeters housed in separate 
cryostats covering the regions up to $|\eta| \approx 4$ (where $\eta$ = 
$-$ln[tan($\theta$/2)] is the pseudorapidity, and $\theta$ is the polar 
angle with respect to the proton beam direction). 
Jet reconstruction uses a cone algorithm \cite{cone} 
with radius ${\cal{R}}=\sqrt{(\Delta\eta)^{2} + (\Delta\phi)^{2}}=0.5$
in pseudorapidity and azimuthal angle ($\phi$) space
about the jet's axis. The jet energy scale was calibrated using the 
transverse energy balance in photon-plus-jet events \cite{scale}.
A muon system outside the calorimeters consists of a layer of drift tubes 
and scintillation counters before 1.8 T iron toroids and two similar layers 
outside the toroids. Identified muons were required to have hits in both the 
wire chambers and scintillation counters and were matched to a central 
track which determined their momenta. The missing transverse energy,
$\etmiss$, was determined by the vector sum of the transverse 
components of the energy deposited in the calorimeter and the $p_T$ 
of detected muons. 

Data collection used a three level trigger system and two 
trigger selections were analyzed for the results presented here. 
The first, called the missing energy trigger here, used  
missing energy plus jets elements. At Level~1 it required at least 
three calorimeter trigger towers with $E_{T} > 5$ GeV, where a trigger 
tower spans $\Delta\phi \times \Delta\eta=0.2\times 0.2$. The vector 
sum of all jets' transverse momenta, 
defined as $\htmiss \equiv |\sum_{\text{jets}}\vec{p_{t}}|$, was 
required to be greater than $ 20$~GeV at Level~2 and greater than 
$ 30$~GeV at Level~3. For 16\% of the integrated luminosity, the acoplanarity, 
defined as the azimuthal angle between the two leading jets, 
was required to be less than $169^\circ$ and the 
$H_T \equiv \sum_{\text{jets}}|\vec{p_{t}}|$ be greater than $ 50$~GeV.
An integrated luminosity of 360~pb$^{-1}$ \cite{newlum} was collected with 
this trigger. The second trigger, called the muon trigger here, used muon and jet 
elements to increase the acceptance for events where one of the $b$ jets 
was identified by its associated muon. At Level~1 it required at least one muon
candidate and at least one calorimeter trigger tower with $E_{T} > 3$ GeV. 
Higher jet thresholds were imposed at Level~2
and finally $25$~GeV at Level~3. An integrated luminosity of 
425~pb$^{-1}$ was collected with the muon trigger.
These missing energy and muon triggers were not independent and only the 
65~pb$^{-1}$ of the muon trigger data 
sample which does not overlap was used for the combined result.
\begin{table*}[t]
\caption{\label{table-mht-mujet-summary}Predicted numbers of 
signal and background events  
  before $b$ tagging and after all requirements (statistical errors 
only).}
\begin{ruledtabular}
\begin{tabular}{lcccc}
Data sample & \multicolumn{2}{c}{Missing energy trigger 360 pb$^{-1}$} 
& 
\multicolumn{2}{c}{Muon 
trigger 425 pb$^{-1}$} \\
\hline
       Process   &  Pretag requirements & All requirements   & Pretag 
requirements & All requirements  \\
     \hline
$W~\rightarrow~\mu\nu~+~jj$         &$\enspace108~\pm~6~$~&~$0.28~\pm~0.11~~$&~$~100~\pm~7~~~~$~&~$0.06~\pm~0.06~$~\\
$W~\rightarrow~e\nu~+~jj$~          &~~$160~\pm~14$~&~$0.02~\pm~0.01~$~      &~$~~~6~\pm~3~~~$~&~$~~~~~0~~~~~~$~\\
$W~\rightarrow~\tau\nu+jj$~         &~~$396~\pm~36$~&~$0.17~\pm~0.05~$~      &~$~~~7~\pm~5~~~$~&~$~~~~~0~~~~~~$~\\
$Z~\rightarrow~\nu\bar{\nu}~+~jj$~  &~~$603~\pm~18$~&~$0.45~\pm~0.16~$~      &~$\enspace25~\pm~4~~~$~&~$~~~~~0~~~~~~$~\\
$t\bar{t}$ and single top           &~~$~36~\pm~1~$~&~$1.42~\pm~0.11~$~      &~$~~18~\pm~0.6~$~&~$0.80~\pm~0.11~$~\\
$W/Z + c\bar{c}$                    &~~$~18~\pm~1~$~&~$0.46~\pm~0.11~$~      &~$3.01~\pm~0.49$~&~$0.21~\pm~0.12~$~\\
$Z + b\bar{b}$                      &~~$~~6~\pm~1~$~&~$0.67~\pm~0.08~$~      &~$1.89~\pm~0.20$~&~$0.22~\pm~0.06~$~\\
$W + b\bar{b}$                      &~~$~~8~\pm~1~$~&~$0.59~\pm~0.11~$~      &~$4.43~\pm~0.38$~&~$0.41~\pm~0.11~$~\\
      \hline
Total SM expected                   &$1335~\pm~43$~&~~$4.1~\pm~0.3~~$~       &~$~165~\pm~10~~~$~&~$~1.7~\pm~0.2~~$~\\
QCD contribution                    &~$~~40~\pm~40$~&~~~~~$~<~0.1~~$~        &~$~~~6~\pm~6~~~$~&~$~~~~~<~0.2~~$~\\
     \hline
     Data                      &    ~\enspace1241&   1            & 146           & 0             \\
    \hline
Signal $M_{LQ}=200$~GeV        &~~$~34~\pm~1~~$~&~$10.1~\pm~0.3~~$~&~$~~9.6~\pm~0.4~~$~
&~$~3.8~\pm~0.2~~$~\\
Signal acceptance              &~~~~~~~~35.9\%~~~~~&~~10.4\% &~~~8.4\%&~~3.3\% \\
\end{tabular}
\end{ruledtabular}
\end{table*}

Signal samples for leptoquark masses between 150 and 400~GeV were 
generated with {\sc pythia} 6.202 \cite{Pythia}. Instrumental background 
comes mostly from QCD multijet processes with false $\etmiss$ arising from 
mismeasurement, and dominates the low $\etmiss$ region. Physics 
backgrounds are SM processes with real $\etmiss$ and were 
estimated from Monte Carlo (MC) simulations. The most important are 
leptonic decays of $W/Z$ bosons plus jets with $Z~\rightarrow~\nu\bar{\nu}$ 
or when a lepton remains unidentified or is misidentified 
as a hadron, and processes which produce top quarks. For all MC 
samples except $t\bar{t}$ and single top quark, the 
next-to-leading order cross sections were obtained from Ref.~\cite{mc_xsec}.
Cross sections for $t\bar{t}$ and single top quark production were taken 
from Ref.~\cite{topD0} and \cite{singletop}, respectively.
At the parton level,  single top quark MC events were generated with 
{\sc comphep} 4.4 \cite{CompHEP}, and  {\sc alpgen} \cite{Alpgen} was 
used for all other samples. These events were then processed with {\sc pythia} 
which performed showering and hadronization. An average of 0.8 minimum bias 
events was superimposed on each MC event to match the number of additional collisions 
observed in data. The resulting samples were processed using a full {\sc geant} 
simulation of the D0 detector \cite{GEANT}. {\sc cteq5l} 
\cite{PDF:CTEQ5L} was used as the parton density function in all cases.

For both data samples, a set of preselection requirements 
was applied prior to $b$ tagging in order to
reduce the number of events from QCD multijet and $W/Z$+jets processes. 
Values for preselection cuts and jet quality criteria were driven 
by trigger requirements. To reject $W\rightarrow \ell\nu$ decays, a veto 
was applied to events with isolated electrons or muons with $p_T>5$ GeV. 
Likewise, events containing
a track with tighter isolation cuts and with $p_T>5$ GeV were rejected 
to reduce the contribution of leptons which remained unidentified. 
The number of events with mismeasured $\etmiss$ was reduced by 
requiring that the primary vertex be within 
$\pm$60~cm in the beam direction from the center of the detector 
and by eliminating those where the $\etmiss$ direction and a jet 
overlapped in $\phi$. For the missing energy trigger sample, events were 
required to have $\etmiss >70 $ GeV, the leading jet was required to have 
$\vert\eta\vert<1.5$ and $p_T>40$~GeV, and, for events without muons, 
scalar $H_T > 110$ GeV. For the muon triggered sample, the preselection required 
a muon with $p_{T} > 4$~GeV and a leading jet with $|\eta|< 1.5$ and $p_{T}>40$~GeV 
($>50$ GeV if not associated with a muon). Additional requirements were a 
second jet with $p_{T} > 20$~GeV, $\htmiss > 50$~GeV and $\etmiss>70$~GeV. 
The numbers of pre-selected events in both samples and their 
estimated sources are given in Table~\ref{table-mht-mujet-summary}.

\begin{figure}[hb]
 \leftline{
 \includegraphics[scale=0.21]{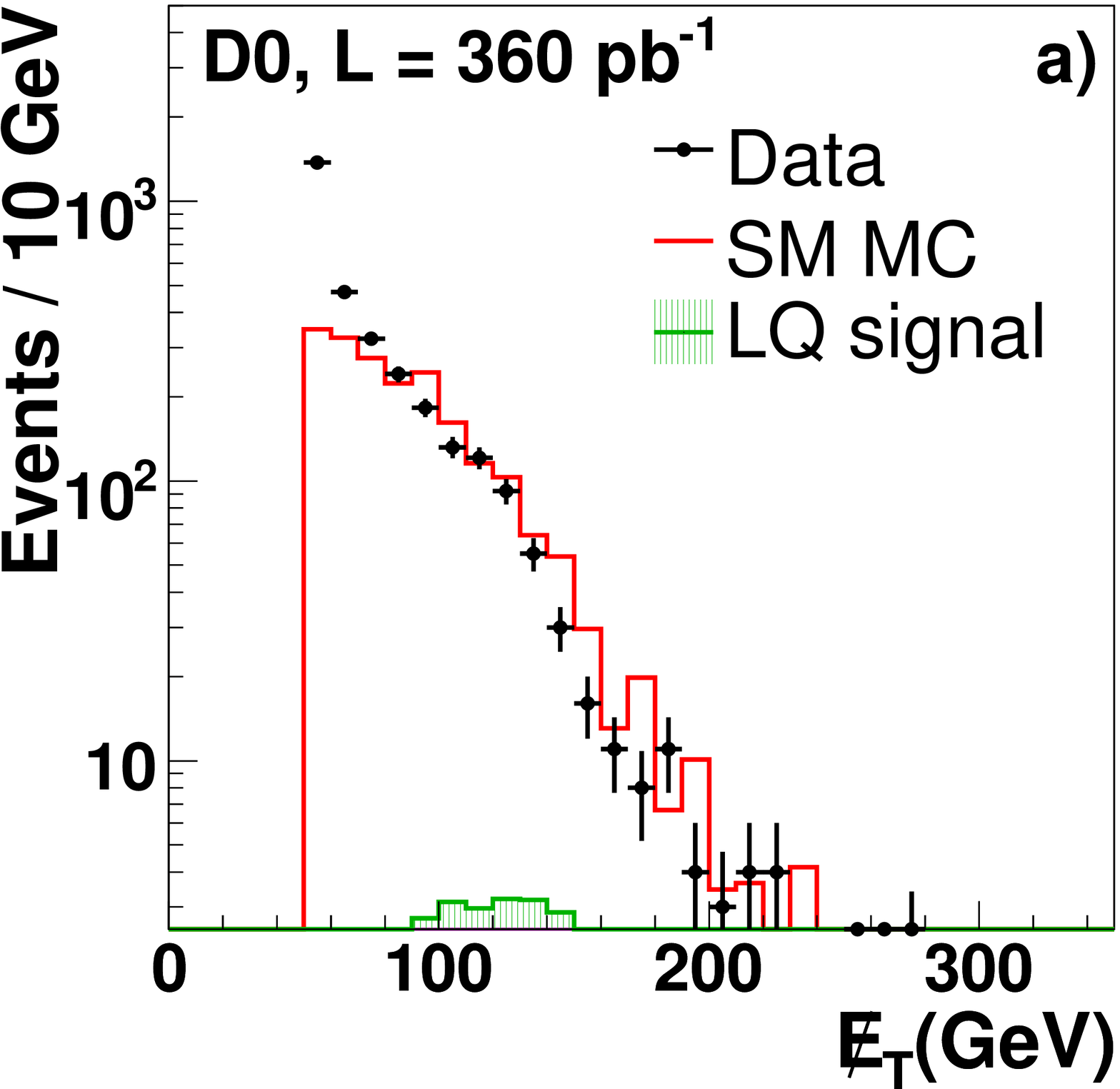} 
 \includegraphics[scale=0.21]{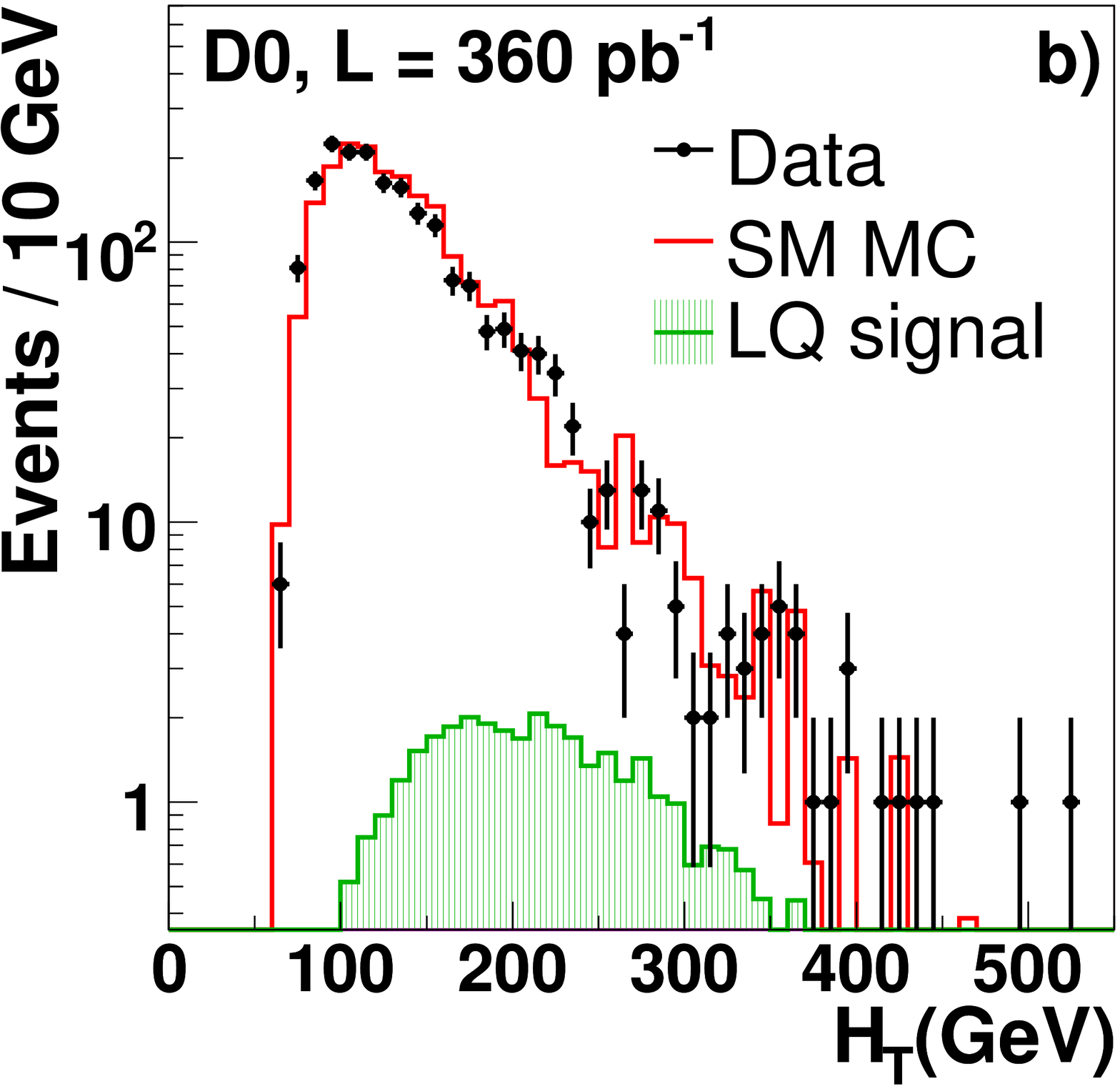}
 }
 \leftline{
 \includegraphics[scale=0.21]{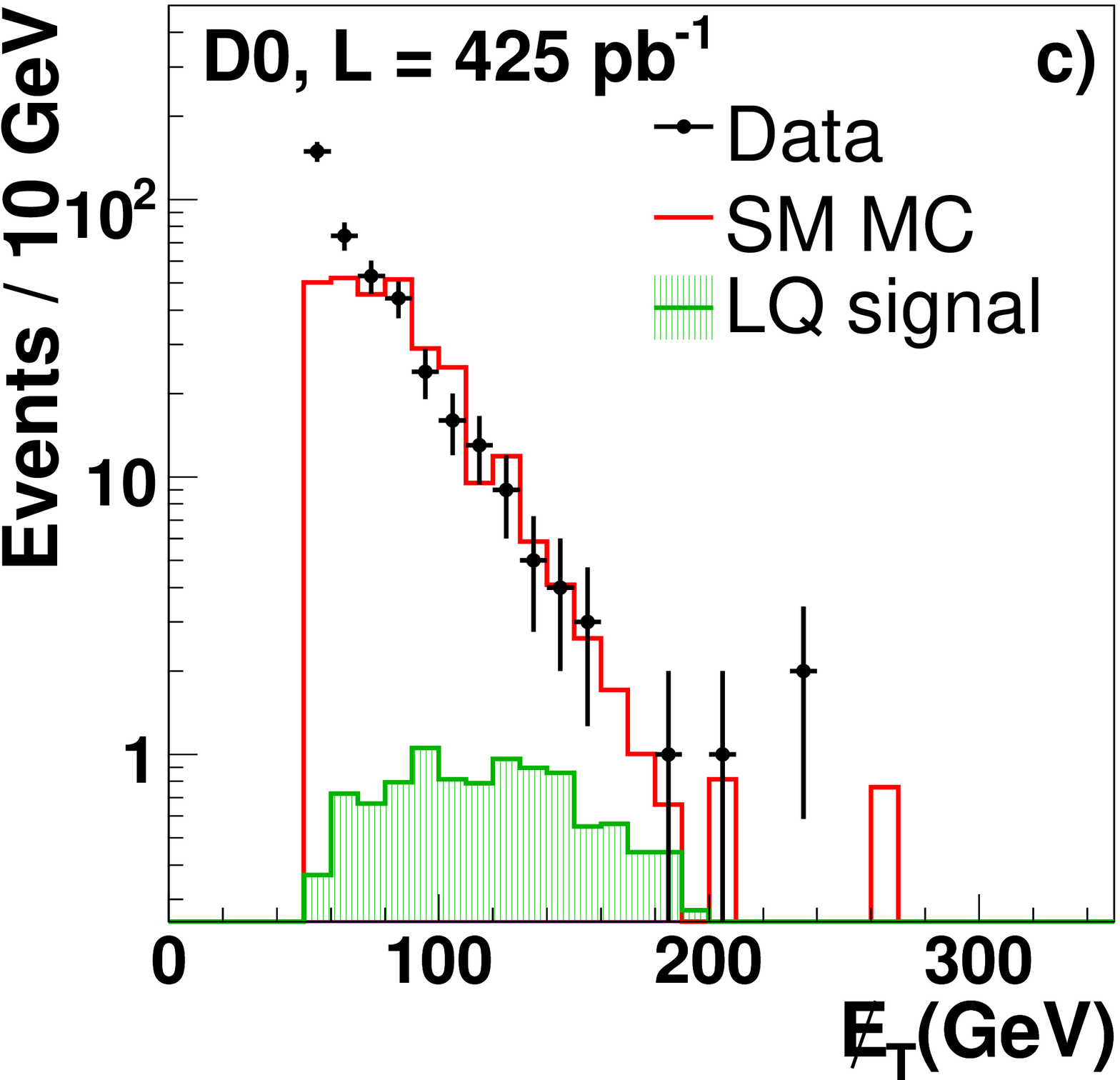} 
 \includegraphics[scale=0.21]{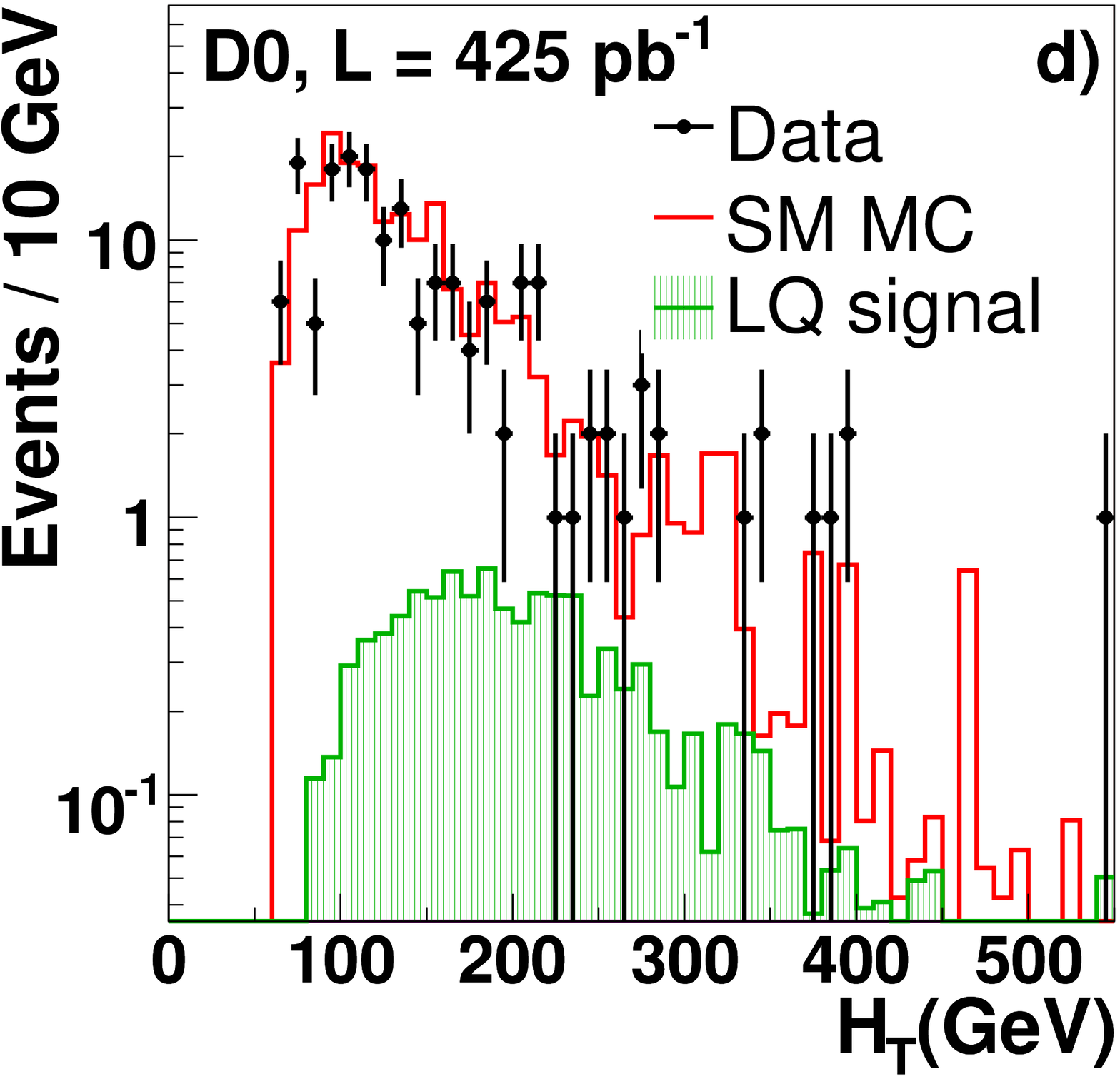}
 }
 \caption{\label{fig:met_and_mht_pretag_np} 
The $\etmiss$ distributions and the scalar $H_T$ (with 
$\etmiss > 70$~GeV) distributions before $b$ tagging for data (points) 
compared to SM background (solid histogram). The missing energy trigger sample is
given in (a) and (b) and the muon trigger sample in (c) and (d).
The shaded histograms are the expected contribution for a 200~GeV LQ signal.}
\end{figure}

\begin{table*}[ht]
\caption{\label{tab:table_95CLcombo_NPcf}
Numbers of observed and predicted events after final selection, the effective 
signal acceptance (with total error), and the observed and expected 95\%~C.L. 
cross section limits as a function of~$M_{LQ}$. Scalar cross sections were used to 
calculate the expected numbers of signal events.}
\begin{ruledtabular}    
\begin{tabular}{lccccccccc}
$M_{LQ}$ & ($\etmiss$, $H_{T}$)\footnote[1]
           {$\etmiss>70$ GeV, $H_{T}>$140 GeV applied to all 
muon-tagged events.} 
&  Data   &  SM $\pm$ stat $\pm$ sys 
& Signal $\pm$ stat $\pm$ sys & Effective  &  $\sigma$ 95\% C.L. limit \\
    GeV     & GeV    &  events     &   events    &  events       
& acceptance (\%)       & obs./exp. (pb) \\
\hline
    170   &  (70,110) &4 & 7.3 $\pm$ 0.4 $\pm$ 1.7 & 27.0 $\pm$ 0.6 $\pm$ 4.6 & 10.4 $\pm$ 1.5 & 0.163/0.232 \\
    200   &  (90,150) &1 & 4.3 $\pm$ 0.3 $\pm$ 1.0 & 10.7 $\pm$ 0.3 $\pm$ 1.7 & 11.1 $\pm$ 1.6 & 0.101/0.163 \\
    220   &  (90,190) &1 & 3.3 $\pm$ 0.3 $\pm$ 0.7 & ~5.8 $\pm$ 0.2 
$\pm$ 0.9  & 11.5 $\pm$ 1.6 & 0.097/0.142 \\
    240   &  (90,190) &1 & 3.3 $\pm$ 0.3 $\pm$ 0.7 & ~3.7 $\pm$ 0.1 
$\pm$ 0.6  & 13.6 $\pm$ 2.0 & 0.081/0.119 \\
    280   &  (90,190) &1 & 3.3 $\pm$ 0.3 $\pm$ 0.7 & ~1.3 $\pm$ 0.0 
$\pm$ 0.2  & 15.5 $\pm$ 2.2 & 0.071/0.105 \\
    320   &  (90,190) &1 & 3.3 $\pm$ 0.3 $\pm$ 0.7 &  ---  & 17.5 $\pm$ 2.5 & 0.063/0.092 \\
    360   &  (90,190) &1 & 3.3 $\pm$ 0.3 $\pm$ 0.7 &  ---  & 18.9 $\pm$ 2.7 & 0.058/0.085 \\
    400   &  (90,190) &1 & 3.3 $\pm$ 0.3 $\pm$ 0.7 &  --- & 21.6 $\pm$ 3.1 & 0.051/0.074 \\
\end{tabular}
\end{ruledtabular} 
\end{table*}

Figure~\ref{fig:met_and_mht_pretag_np} 
shows distributions of $\etmiss$ and $H_T$ with the signal LQ and 
background SM events normalized to the total integrated luminosity. 
The data samples reproduce the SM expectations for $\etmiss > 90$~GeV  
indicating that contributions from QCD multijet processes are small in this range.
The contribution from these events is estimated 
from the $\etmiss$ distribution below 70 GeV by a fit to an exponential
after subtracting SM contributions. This is similar to the technique used in our search 
for scalar bottom quarks \cite{sbottom} and 
total, for $\etmiss > 70$~GeV,  $40\pm 40$ events and 
$6\pm 6$ events in the missing energy and muon 
trigger samples, respectively. After $b$ tagging, which is described below, 
the contributions from this source are less than 0.1 and 0.2 events 
respectively, and a value of 0 events was conservatively used for limit calculations.

Backgrounds with light flavor jets were reduced by requiring 
the presence of $b$-tagged jets. We used jets that contained either 
tracks with a significant impact parameter or muons to
select $b$-jet candidates. Events were required to have two $b$ tags 
with at least one  passing the impact parameter criterion.
For events selected with the muon trigger, a $b$ jet tagged
using a reconstructed muon in proximity to a jet was required. Otherwise, 
the events from both trigger samples
were treated in an identical way for the remainder of the analysis. 

We assigned a $b$ probability to a jet based on properties such as the 
existence of tracks with a significant impact paramater that indicated the 
presence of a secondary vertex. The algorithm \cite{clement} required at 
least two tracks in a jet, each with a hit in the silicon tracker.
Tagging probabilities in simulated jets used parameterizations derived from data. 
The probability of a jet to be of light flavor was derived and required to be 
less than 2\%, which yielded a $b$-tag efficiency of about 45\% per $b$ jet. 
This choice maximized the expected LQ mass limits after all other cuts were applied. 

Muon-tagged jets were also considered $b$-jet candidates. Muon thresholds
were raised to $p^{\mu}_{T} >6$ GeV to suppress contributions from 
$\pi/K$ decays. Remaining backgrounds from $W$ boson decays to muons were due to 
accidental overlap of a muon with a nearby jet. We required that 
the sum of track $p_{T}$ in a cone of 0.5 around the muon be greater 
than  10~GeV, and that the approximate $p_T$ of the muon relative to the jet's axis,
$\Delta{\cal{R}}_{{\mu}-\text{jet}} \times p^{\mu}_{T}$, be less than 3.5 GeV,
as muons originating from jets are closer to the jet axis for
higher values of $p_{T}$  \cite{muon_jet_angle}.
These requirements are not independent and combining them  
was found to reduce the $W$ boson background by 95\% while keeping 77\% of the signal. 
Muon tagging has a $b$-tag efficiency of about 11\% with less than	 
0.5\% of light flavored jets passing the tag criteria.

Since signal events are dominated by high energy $b$ jets, the 
quantity $X_{jj} \equiv  ( p_{T}^{\text{tag1}} +  p_{T}^{\text{tag2}} ) / 
(\Sigma_{\text{jets}} p_T)$
was defined, with the muon $p_T$ included in the $p_T$ of the tagged
jet, where applicable. We required $X_{jj}>$ 0.8 which was found to 
significantly reduce the contribution from top quark pair events.
Since $\etmiss$ and $H_T$ increase for higher values of $M_{LQ}$, we
optimized the requirements on these parameters as a function of leptoquark 
mass by maximizing $S/\sqrt{B}$, where $S$ and $B$ are estimated signal and
background rates. The values used for the minimum $H_T$ and $\etmiss$ are 
given in  Table~\ref{tab:table_95CLcombo_NPcf} and were applied only to 
the double $b$ vertex tagged sample. For the muon-tagged events, the $H_T> 140$~GeV 
requirement was applied, and the $\etmiss$ cut remained at 70~GeV 
as these events have a smaller contribution from light flavor jets.

Results of the final event selection along with predicted numbers for 
signal ($M_{LQ}=200$ GeV) and SM backgrounds are listed in 
Table~\ref{table-mht-mujet-summary}. The latter originate mostly from
$W/Z$ + $b\bar{b}$ production and top quark events.

Sources of systematic uncertainties include errors
in the determination of the integrated luminosity (6.1\%) \cite{newlum} 
and SM cross sections (15\%).
Trigger and jet selection efficiencies were measured with data and their
contribution to the systematic errors is small. Jet energies and $\etmiss$
were varied within the energy scale correction uncertainty, and the impact 
on signal acceptance and background rates was determined with MC to
be 3\% and 10\% respectively. Jet $b$-tagging efficiency uncertainties 
are 12\% for signal and 11\% for background.
\begin{figure}[ht]
  \centering
   \includegraphics[scale=.45]{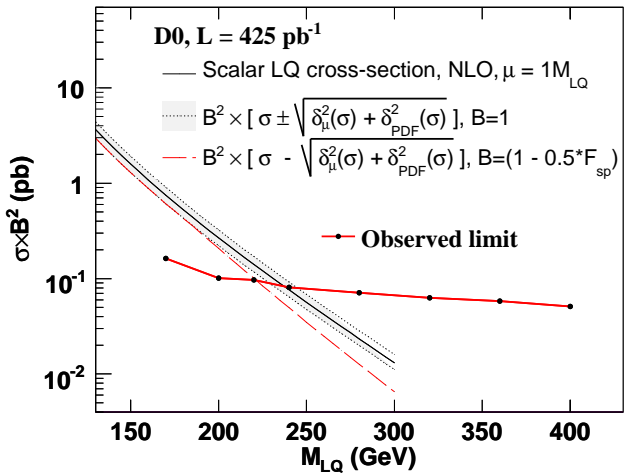}
   \caption{\label{fig:Lq3_Exclusion_final}
 The 95\%~C.L. limit on $\sigma \times B^2$ (points plus
 solid line) as a function of $M_{LQ}$. The 
prediction for scalar leptoquarks (solid line) include an error range (in grey)
of $\mu$ between ${0.5M_{LQ}}$ and $2M_{LQ} $.
The long-dashed line below the theory band indicates 
the threshold effect for the $\tau t$ channel. 
 }
\end{figure}

 One event remains in the combined data sample for the 
selection criteria used for all points with $M_{LQ}\ge$~200~GeV. 
This is consistent with the 3.3~$\pm$~0.3~$\pm$~0.7
expected events from SM processes. The probability of the observed deficit
is 16\%. The 95\% C.L. upper limits on the 
$\sigma(p\overline{p}\rightarrow LQ\overline{LQ} \rightarrow \nu\bar{\nu} 
b\overline{b})\times B^2$ were obtained using the techniques in Ref.~\cite{Junk}. 
The effective signal acceptances of the combined sample (normalized to 
360~pb$^{-1}$), numbers of events, and the resulting limits as  functions 
of~$M_{LQ}$~are~summarized in Table~\ref{tab:table_95CLcombo_NPcf}.

Figure~\ref{fig:Lq3_Exclusion_final} shows the cross section limit as a
function of $M_{LQ}$. Limits on the scalar leptoquark 
mass were obtained by the intersections of the observed 95\%~C.L. cross section 
limits with the lower bounds of a next-to-leading order calculation for which 
variation of the renormalization 
scale $\mu$ from $0.5M_{LQ}$ to $2M_{LQ}$ and the PDF uncertainties
\cite{pdf} were included. If $B(LQ \rightarrow \nu_\tau b)=1$ is assumed, 
our limit is $M_{LQ}>229$~GeV. We can also consider the case where
$LQ\rightarrow t\tau$ decays occur. If we assume that the
leptoquark couplings to $\nu_\tau b$ and $t\tau$ are the same, the 
branching fraction for $LQ \rightarrow \nu_\tau b$ is then $1-0.5\times F_{sp}$ 
where $F_{sp}$ is the phase space suppression factor for the $t\tau$ channel
\footnote[1]{We used $B(LQ \rightarrow \nu b$) = $1 - 0.5\times F_{sp}$ , 
where $F_{sp}=\sqrt{(1+d_{1}-d_{2})^2-4d_{1}}[1-d_{1}-d_{2}]$,  
with $d_{1} = {(m_{t}/M_{LQ})}^2$ and $d_{2} = 
{(m_{\tau}/M_{LQ})}^2$. T. Rizzo (private communication).}.
This is shown on the figure as a displacement from the lower edge of the 
theory band. With this assumption, the 95\%~C.L. lower mass limit for 
scalar~leptoquarks~is~221~GeV. 

In conclusion, we observe one event with the topology $b\bar{b}+\etmiss$ 
consistent with that expected from top quark and $W$ and $Z$ boson 
production and set limits on the cross section times branching fraction squared
to the $b\nu$ final state as a function of leptoquark mass for  
charge-1/3 leptoquarks. These limits are interpreted as mass
limits for third-generation scalar leptoquarks and increase the excluded 
value by 81 GeV compared to previous results.
%
\begin{acknowledgments}
%
We thank the staffs at Fermilab and collaborating institutions, 
and acknowledge support from the 
DOE and NSF (USA);
CEA and CNRS/IN2P3 (France);
FASI, Rosatom and RFBR (Russia);
CAPES, CNPq, FAPERJ, FAPESP and FUNDUNESP (Brazil);
DAE and DST (India);
Colciencias (Colombia);
CONACyT (Mexico);
KRF and KOSEF (Korea);
CONICET and UBACyT (Argentina);
FOM (The Netherlands);
Science and Technology Facilities Council (United Kingdom);
MSMT and GACR (Czech Republic);
CRC Program, CFI, NSERC and WestGrid Project (Canada);
BMBF and DFG (Germany);
SFI (Ireland);
The Swedish Research Council (Sweden);
CAS and CNSF (China);
Alexander von Humboldt Foundation;
and the Marie Curie Program.
%

\end{acknowledgments}

\end{document}